\documentclass[reprint,onecolumn,amsmath,amssymb,jap,aip]{revtex4-2}
\usepackage{xcolor}
\usepackage{units}
\usepackage[T1]{fontenc}
\usepackage{amsmath,amssymb,amsfonts}
\usepackage{units}
\usepackage{algorithmic}
\usepackage{graphicx}
\usepackage{textcomp}
\usepackage{xcolor}
\usepackage{svg}
\usepackage{caption}
\usepackage{subcaption}
\usepackage[utf8]{inputenc}
\usepackage[english]{babel}
\newcommand{\Vth}{$V_\text{th}$}
\newcommand{\Glass}{Glass($t$,$T_\text{hist}$)}

\newcommand{\beginsupplement}{%
	\setcounter{table}{0}
	\renewcommand{\thetable}{S\arabic{table}}%
	\setcounter{figure}{0}
	\renewcommand{\thefigure}{S\arabic{figure}}%
	\setcounter{section}{0}
	\renewcommand{\thesection}{S\arabic{section}}%
}

\begin{document}

\title{Measurement of onset of structural relaxation in melt-quenched phase change materials}

\author{Benedikt Kersting} 
\affiliation{IBM Research - Europe, S\"{a}umerstrasse 4, 8803 R\"{u}schlikon, Switzerland.}
\affiliation{Westf\"{a}lische Wilhelms Universit\"{a}t M\"{u}nster, Institute of Materials Physics, Wilhelm-Klemm-Straße 10, 48149 M\"{u}nster, Germany.}

\author{Syed Ghazi Sarwat }
\affiliation{IBM Research - Europe, S\"{a}umerstrasse 4, 8803 R\"{u}schlikon, Switzerland.}

\author{Manuel Le Gallo}
\affiliation{IBM Research - Europe, S\"{a}umerstrasse 4, 8803 R\"{u}schlikon, Switzerland.}

\author{Kevin Brew}
\affiliation{IBM Research AI Hardware Center, 257 Fuller Road, Albany, NY, US.}

\author{Sebastian Walfort }
\affiliation{Westf\"{a}lische Wilhelms Universit\"{a}t M\"{u}nster, Institute of Materials Physics, Wilhelm-Klemm-Straße 10, 48149 M\"{u}nster, Germany.}

\author{Nicole Saulnier}
\affiliation{IBM Research AI Hardware Center, 257 Fuller Road, Albany, NY, US.}

\author{Martin Salinga }
\affiliation{Westf\"{a}lische Wilhelms Universit\"{a}t M\"{u}nster, Institute of Materials Physics, Wilhelm-Klemm-Straße 10, 48149 M\"{u}nster, Germany.}

\author{Abu Sebastian}
\affiliation{IBM Research - Europe, S\"{a}umerstrasse 4, 8803 R\"{u}schlikon, Switzerland.}

\date{\today}
\begin{abstract}

{Chalcogenide phase change materials enable non-volatile, low-latency storage-class memory. They are also being explored for new forms of computing such as neuromorphic and in-memory computing. A key challenge, however, is the temporal drift in the electrical resistance of the amorphous states that encode data. Drift, caused by the spontaneous structural relaxation of the newly recreated melt-quenched amorphous phase, has consistently been observed to have a logarithmic dependence in time. Here, we show that this observation is valid only in a certain observable timescale. Using threshold-switching voltage as the measured variable, based on temperature-dependent and short timescale electrical characterization, we experimentally measure the onset of drift. This additional feature of the structural relaxation dynamics serves as a new benchmark to appraise the different classical models to explain drift.} 

\end{abstract}

\maketitle


\section{Introduction}

\begin{figure}
 \centering
 \includegraphics[width=0.75\linewidth]{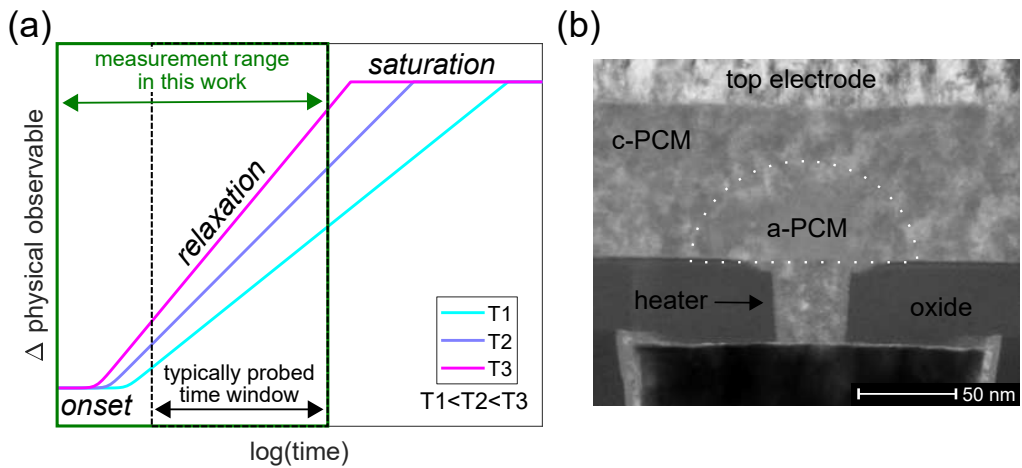}
 \caption{(a) Sketch of the temporal evolution of material properties upon structural relaxation: Previous studies of phase change materials have been limited to time scales and temperatures in which drift obeys a $\log(t)$ dependence. In this work, we expand the measurement range to the onset of relaxation and study its temperature dependence. Since relaxation is thermally activated, the onset and saturation of drift shift to shorter time scales with increasing ambient temperature. (b) Transmission electron micrograph of the mushroom-type PCM device used in this study. The amorphous dome that is created by passing a current through the narrow bottom electrode is highlighted with a white dotted line.}
 \label{fig:SketchSigmoidRelaxation}
\end{figure}

\noindent When a liquid is quenched faster than its critical cooling rate, crystallization events can be overcome and atoms can structurally freeze into a disordered solid state.\textsuperscript{\cite{Turnbull1969,Zhong2014a,Salinga2018}} Upon cooling the melt, the atomic mobility decreases. Eventually, the system can no longer assume the equilibrium structure of the supercooled liquid state within the timescale of the experiment and a non-equilibrium glass state is created. The free energy difference between the super-cooled liquid and glass state results in structural relaxation, where the atomic configurations in the glass change over time. The intrinsic material properties, including viscosity, density, and the electronic bandgap, change due to relaxation,\textsuperscript{\cite{Angell2000b,Priestley2005a}} and this is understood to occur in three phases (Figure~\ref{fig:SketchSigmoidRelaxation}a).\textsuperscript{\cite{Chen2007,McKenna2003,Priestley2009}} For every rearrangement, a finite energy barrier must be overcome, and therefore, for some amount of time, the onset phase, the properties do not change. In the second phase, where relaxation is most profound, the properties have been observed to change proportionally to log(t). Finally, approaching the supercooled liquid, the glass reaches a saturation phase, and the properties no longer continue to change. Tracking this structural relaxation process through all three phases is experimentally challenging. Most studies are focused on amorphous polymers and metallic glasses, yet there is little work on highly fragile glass formers with bad glass forming ability, such as phase change chalcogenide glasses. 

Thin films of phase change chalcogenides, such as Ge$_2$Sb$_2$Te$_5$ (GST) show interesting electrical and optical material properties, which can be rendered tunable via rapid and reversible crystalline to amorphous phase-transitions. Phase change chalcogenides are exploited for many technologies, including the commercialized electrical phase change memory (PCM) technology. In PCM, a nanoscale volume of a chalcogenide compound is sandwiched between the top and bottom electrodes. Joule heating, from current across the electrodes (Figure~\ref{fig:SketchSigmoidRelaxation}b), allows reversible amorphization and crystallization of the chalcogenide glass.\textsuperscript{\cite{LeGallo2020AnPhysics}} The amorphous and crystalline states exhibit electrically distinct properties and thus, the device resistance can be toggled between a electrically conductive state (SET) and resistive state (RESET). Within PCM devices, the amorphous state relaxes after RESET, and the observable metrics, such as the electrical resistance ($R$) and the threshold-switching voltage (\Vth), change due to structural relaxation. This process is commonly referred to as drift.\textsuperscript{\cite{Zhang2020a}} 

In the RESET state, the device resistance increases with $\log(R) = \log(R_0) + \nu_\text{R} * \log(t/t_0)$, and the threshold-switching voltage increases with $V_\text{th} = V_\text{th,0} + \nu_\text{Vth} * \log(t/t_0)$, where $\nu_\text{R}$ and $\nu_\text{th}$ denotes the resistance and threshold voltage drift coefficients, respectively, $R_0$ the resistance at $t_0$ and $V_{th,0}$ the threshold voltage at $t_0$. Importantly, however, these equations are only valid in the relaxation phase. What remains to be investigated both qualitatively and quantitatively are the other two phases, namely, when does structural relaxation begin and when does it end? To this end, measurements that capture the relaxation from extremely short to long timescales at different temperatures are required. Although drift in PCM devices has been studied extensively experiments only revealed the $\log(t)$ dependency of resistance and threshold voltage.\textsuperscript{\cite{Gorbenko2019,LeGallo2018,Salinga2018}} A notable exception is a stand-alone sub-\unit[100]{$ns$} drift measurement by Ielmini et al. on GST that hints at the presence of a region where drift is absent.\textsuperscript{\cite{Ielmini2007}} In Supplementary Note 1 we compare these experiments to our study. 

The goal of this study is to shed light on the phase where relaxation is absent. Specifically, to quantify on what timescales the commonly assumed $\log(t)$ dependence is valid, and to appraise the different classical models to explain drift. To this end, we employ \Vth~as a means to observe the state of relaxation. We study the drift characteristics of GST and a doped GST (dGST) by setting up a \Vth~drift measurement and analysis framework. Mushroom-type PCM devices of both materials are melt-quenched at temperatures spanning from \unit[100]{$K$} to \unit[300]{$K$}, and drift is probed from tens of nanoseconds to ten seconds after RESET. The experimental data are fitted with two models, namely the collective relaxation and the Gibbs model of relaxation, and the different physical parameters used in the fitting are discussed and compared. 

\section{Threshold-switching voltage drift experiments}\label{sec:expdata}

\begin{figure}
 \centering
 \includegraphics[width=0.9\linewidth]{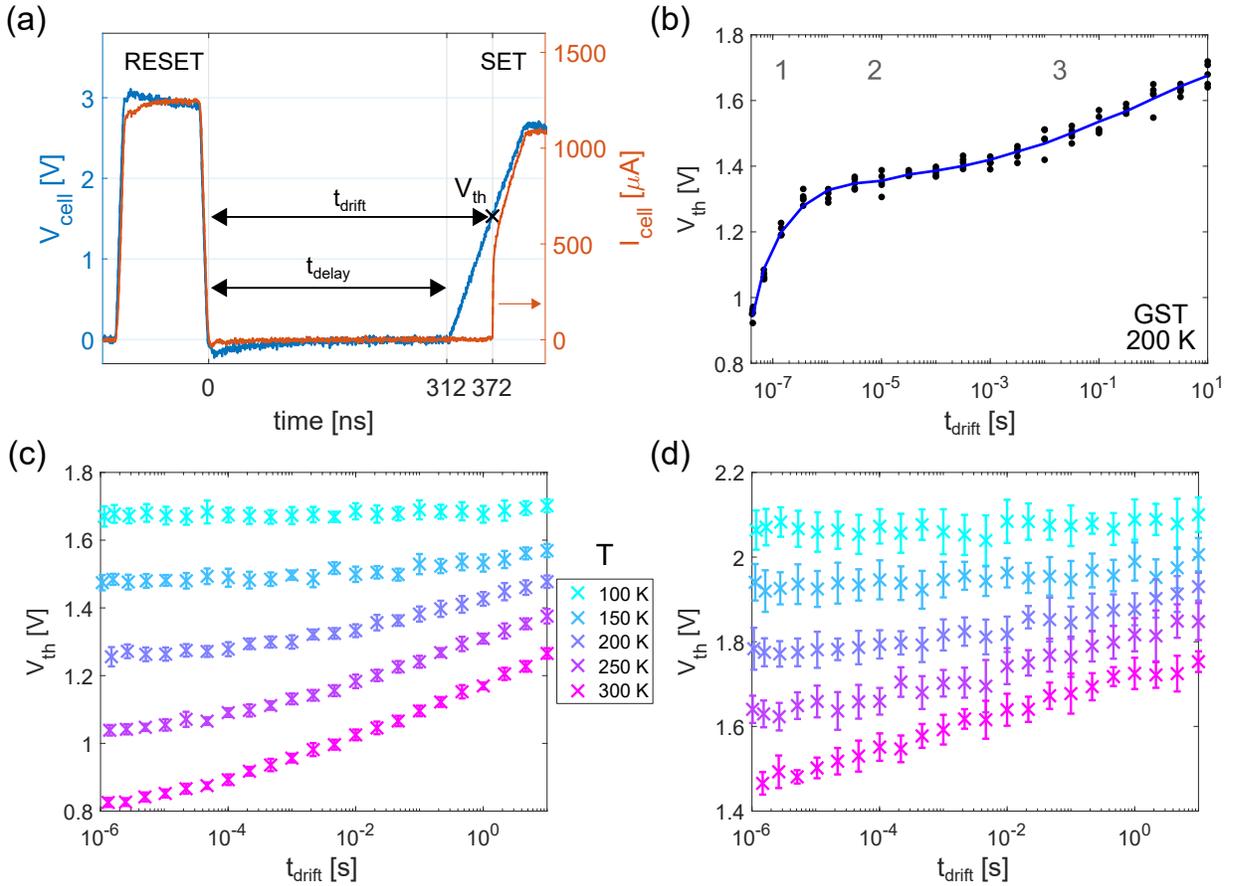}
 \caption{Threshold-switching voltage drift experiment: (a) Programming scheme. A RESET pulse with a \unit[3]{$ns$} falling edge programs the device to a melt-quenched amorphous state. A SET pulse with a \unit[100]{$ns$} leading edge is applied to probe the threshold voltage value. In the example shown here, the device drifts for \unit[372]{$ns$} before threshold switching occurs at \unit[1.5]{$V$}. The time interval (t\textsubscript{delay}) between the RESET and SET pulses is varied to probe the temporal evolution. (b) \Vth~drift in GST. The \Vth~evolution exhibits three distinct regimes (labeled in grey). First a steep increase up to \unit[$\sim$1]{$\mu s$}, second, a flattening to almost constant values, and third a transition to a monotonic increase approximately proportional to log(t). (c) Temperature-dependent \Vth~drift of GST and (d) doped GST in regimes 2 and 3. With increasing ambient temperature, \Vth~begins to change at shorter timescales and shows a larger change with time. Error bars show the standard deviation over 15 measurements.} 
 \label{fig:SummaryExperiment}
\end{figure}

Because structural relaxation processes are thermally activated, ambient temperature can be used as a knob to shift the onset and saturation of drift to experimentally accessible timescales (see Figure~\ref{fig:SketchSigmoidRelaxation}a). However, the observation of the saturation phase by raising the ambient temperature (greater than \unit[400]{$K$}) is prohibitive due to potential recrystallization of the amorphous phase. On the other hand, there is a potential for measuring the onset of drift by monitoring it at lower ambient temperatures. The challenge however, is the inability to reliably measure electrical resistance at short-timescales and low temperatures. Hence, we resort to \Vth~as a means to observe the state of relaxation. \Vth~marks the switching of the highly resistive RESET state to an electronically excited on-state (see Figure~\ref{fig:SummaryExperiment}a).

To probe the \Vth~drift, the mushroom cell is repeatedly programmed to a new RESET state and SET pulses with delay times varying from \unit[10]{$ns$} to \unit[10]{$s$} are applied. Note that each \Vth~measurement results in the erasure by recrystallization of the corresponding RESET state. The measured \Vth~drift represents an averaged behavior of RESET states created in the device. Details on the experimental protocol and the algorithm to define \Vth~are provided in the Methods. Three distinct regimes are apparent in the temporal evolution (Figure~\ref{fig:SummaryExperiment}b). In regime 1, up to $\sim 1 \: \mu s$ there is a steep increase of \Vth. Most likely this is caused by the decay of the RESET excitation. While previous studies attributed this regime solely to decay of the electrical excitation,\textsuperscript{\cite{Elliott2020,Ielmini2007}} thermal transient effects may also play an important role. The threshold voltage increase with time in regime 1 appears to be independent of the ambient temperature (Supplementary Note 2). Regime 2 shows a flattening of the curve and almost constant threshold voltage values. Finally, in regime 3 we observe a continuous linear increase with $\log(t)$. We attribute the temporal evolution in regimes 2 and 3 to the structural relaxation of the amorphous phase with the transition between them marking the onset of relaxation. 


In the following, we further analyze the temperature dependence of regimes 2 and 3 for GST and dGST. To create comparable RESET states at different ambient temperatures, the programming power was scaled such that the initially molten volume remains approximately constant (Supplementary Note 3) and the RESET pulse trailing edge was kept constant. Both materials show the same characteristic behavior (Figure \ref{fig:SummaryExperiment}c and \ref{fig:SummaryExperiment}d). With increasing ambient temperature, the \Vth~values decrease. This is due to the thermally activated transport of the amorphous phase. All experiments capture regime 2 in which \Vth~hardly changes. The transition point to regime 3, the onset of relaxation, shifts continuously to shorter timescales. Furthermore, the slope in regime 3 progressively increases with increasing temperature. Both these effects, namely, the onset shift and the slope change are expected because the relaxation processes are accelerated by increasing temperature. 
At \unit[100]{$K$}, we find \Vth~drift to be absent, which can be either because the drift coefficient is very small or because the onset has shifted outside of the measurement range. One possible reason for the former is that in a phase change material with trap states deep within the band gap, like Ge$_2$Sb$_2$Te$_5$ \textsuperscript{\cite{Luckas2010,Rutten2019,Konstantinou2019}}, resistance drift at very low temperatures may not be observable if the electrical transport changes from a trap limited band transport to a hopping type transport. The activation energy for hopping would be defined by the distance between the Fermi level and the trap states, which may not necessarily change upon structural relaxation. 

\section{Analytical Framework and Relaxation Models}\label{sec:analysis}

In this section, we will try to justify the use of \Vth~to monitor the state of relaxation and also to analyze the experimental data presented in Section \ref{sec:expdata} based on state-of-the-art relaxation models. Establishing an analytical relation between \Vth~and the state of relaxation, \Glass, where $T_{hist}$ captures the thermal history, poses a non-trivial problem. First, the exact mechanism of threshold switching is still debated, and second, it is not clear which material parameters change with relaxation. 

We assume that \Vth~can be defined by the sum of a temperature-dependent function $f(T)$, a term proportional to \Glass~and an offset value ($C_2$) that could change with the size of the amorphous dome for example. 

\begin{equation}
\label{equ:VariableSeparation}
 V_\text{th} = f(T) + C_1 * \mathrm{Glass}(t,T_{hist}) + C_2
\end{equation}

A key basis for this assumption is the approximately linear change of $V_\text{th}$ with the activation energy for electrical conduction ($E_\text{a}$).\textsuperscript{\cite{Pirovano2004,LeGallo2016}} $E_\text{a}$ in turn has been shown to increase with drift,\textsuperscript{\cite{Boniardi2011,LeGallo2018,Wimmer2014b}} and a linear increase of $E_\text{a}$ with $\log(t)$ has been experimentally measured for different phase change materials.\textsuperscript{\cite{Fantini2012,Rutten2015}} Thus, we expect that \Vth~is proportional to \Glass. Finally, we assume that the change of \Vth~upon relaxation is decoupled from the change of ambient temperature $T$. Such a decoupling was previously deduced in \cite{Ciocchini2012} and can also be derived from simulations based on a thermally assisted switching model (see Supplementary Note 4).\textsuperscript{\cite{LeGallo2016}}

From Equation \ref{equ:VariableSeparation}, it can be seen that the temporal change in \Vth~with respect to a defined reference point, $t_\text{ref}$, is dependent only on \Glass. For our analysis, we identify \Vth(1$\mu$s) as an ideal reference point where drift is absent for all temperatures studied. We denote the initially created glass state, that did not yet begin to relax, as $\mathrm{Glass_0}$.

\begin{equation}
\label{equ:DeltVthDeltaGlass}
 \Delta V_\text{th} = V_\text{th}(t,T)-V_\text{th}(1 \mu s,T) = C_1 * (\mathrm{Glass}(t,T_{hist})-\mathrm{Glass_0})
\end{equation}

In the following, we will show that, based on these assumptions, the temperature-dependent onset of threshold voltage drift can be captured with two common relaxation models proposed for phase change materials, that is the Gibbs model \textsuperscript{\cite{Ielmini2008a,Lavizzari2009}} and the collective relaxation model.\textsuperscript{\cite{Sebastian2015,LeGallo2018}}

\subsection*{Collective relaxation model}

\noindent The collective relaxation model does not specify individual relaxation processes or defect states but instead quantifies the relaxation state of the glass by an abstract state variable $\Sigma$. $\Sigma = 1$ is an infinitely unrelaxed state and $\Sigma = 0$ denotes that the system approaches equilibrium.\textsuperscript{\cite{Knoll2009}} Upon relaxation, the system assumes configurational states of progressively lower energy. The activation energy $E_\text{b}=E_\text{s}*(1-\Sigma)$ that must be overcome for the next relaxation step increases monotonically. The temporal evolution of the state variable $\Sigma$ is captured by the rate equation 
\begin{equation}
\label{equ:DifferentialSigma}
\frac{d\Sigma(t)}{dt}= - \nu_0 \Delta_\Sigma *\exp\left(\frac{-E_\text{s}*(1-\Sigma(t))}{k_\text{b}T}\right)
\end{equation}
assuming an Arrhenius dependence of the relaxation rate on the activation energy. The attempt to relax frequency, which is on the order of phonon frequencies, is denoted by $\nu_0$ and $\Delta_\Sigma$ is the change of $\Sigma$ with each relaxation step. Consequently, $\Delta_\Sigma*E_\text{s}$ defines the increase of the activation energy for each subsequent relaxation step. For a constant ambient temperature, the differential equation can be solved analytically as
\begin{equation}
\label{equ:SigmaOftime}
\Sigma(t,T) = -\frac{k_\text{b}T}{E_\text{s}} \log\left(\frac{t+\tau_0}{\tau_1}\right) 
\end{equation}
with $\tau_0 = \frac{k_\text{b}T}{\nu_0\Delta_\Sigma E_\text{s}} \exp(\frac{E_\text{s}*(1-\Sigma_0)}{k_\text{b}T})$ marking the begin of relaxation from the initial glass state $\Sigma_0$ and $\tau_1 = \frac{k_\text{b}T}{\nu_0\Delta_\Sigma E_\text{s}} \exp(\frac{E_\text{s}}{k_\text{b}T})$ the time at which the system reaches equilibrium. In the range $\tau_0<<t<\tau_1$ the temporal evolution of $\Sigma$ follows $\log(t)$ and the change of $\Sigma$ depends linearly on the ambient temperature. The linear temperature dependence of $\tau_0$ is almost negligible compared to the exponential term. In a first approximation the shift of the relaxation onset with temperature is defined by the smallest activation energy for relaxation $E_\text{min} = E_\text{s} * (1-\Sigma_0)$ and the effective attempt to relax frequency $\nu_0 \Delta_\Sigma$ is a scaling factor defining the timescale of the onset.

\begin{figure}
 \centering
 \includegraphics[width=1\linewidth]{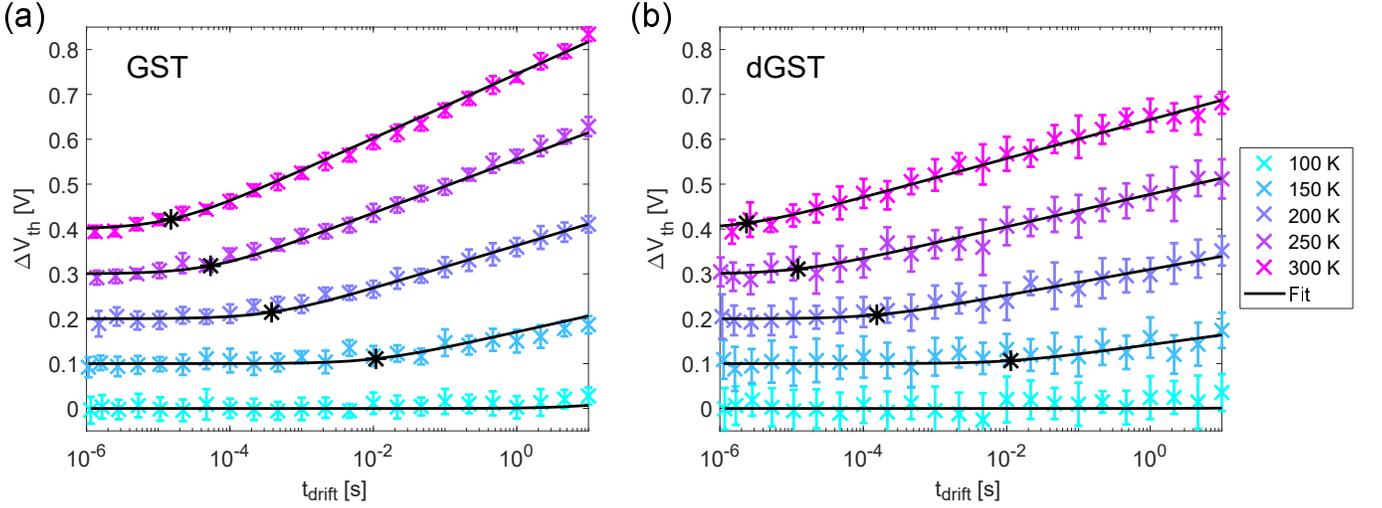}
 \caption{Model fit - Collective relaxation model: The temporal evolution of \Vth~at different temperatures is fitted collectively with Equations (\ref{equ:SigmaOftime}) and (\ref{equ:DeltVthDeltaGlass}). The onset of relaxation is marked with an asterisk. For better visibility, the experiments at different temperatures are shifted along the y-axis by $(T-100 K) * 0.2 V/K$. The fitting parameters are summarized in Table \ref{tab:CollectiveRelaxation}.}
 \label{fig:ModelFit}
\end{figure}

To fit the experimental data with the collective relaxation model, the terms $\mathrm{Glass}(t,T_{hist})$ and $\mathrm{Glass_0}$ in Equation~\ref{equ:DeltVthDeltaGlass} are replaced by $\Sigma(t,T)$ and $\Sigma_0$ respectively. Both distinct features, the shift of the onset $\tau_0$ and the increase of the drift coefficient $\nu_\text{Vth} = C_1 * k_\text{b}T /E_\text{s} $ with ambient temperature are well captured (Figure \ref{fig:ModelFit}). It confirms that the observed threshold voltage evolution is caused by the relaxation dynamics of the amorphous phase. At \unit[300]{$K$}, $\tau_0$ of GST and dGST is \unit[$\sim 15$]{$\mu s$} and \unit[$\sim 2.3$]{$\mu s$} after RESET, respectively. Interestingly, the relaxation onset of the glass states created at ambient temperatures ranging from \unit[100]{$K$} to \unit[300]{$K$} can be fitted with a single $\Sigma_0$, i.e. the degree of relaxation of the initially created glass state does not change notably. Faster quenching has been observed to create less relaxed glass states.\textsuperscript{\cite{Greer1982}} In our study the cooling profile of the melt-quenching process changes with ambient temperature. An understanding of the quench-rates and glass transition temperature in the device and how these determine the initial value of $\Sigma_0$ could be subject of future work.

Still it is not possible to determine unique values of $\Sigma_0$, $C_1$, $E_s$, or $\nu_0 \Delta_\Sigma$, the material parameters defining the relaxation kinetics. To this end it would be necessary to also determine the time of drift saturation $\tau_1$. As long as only the drift coefficient $\nu_{Vth}$ and $\tau_0$ are known, the four fitting variables are interdependent. The drift coefficient depends on $C_1/E_s$, the exponential prefactor of $\tau_0$ on $\nu_0 \Delta_\Sigma E_s$ and the exponential term on $(1-\Sigma_0 ) E_s$. 

\begin{table}[h!]
 \centering
 \begin{tabular}{c c c c}
 \hline
 Material & $C_1 /E_s$ [V/eV] & $\nu_0 \Delta_\Sigma E_s$ [eV/s] & $(1-\Sigma_0)E_s$ [eV] \\
 \hline
 GST & -1.2 & 2.48e6 & 0.19 \\ 
 dGST & -0.73 & 1.07e8 & 0.24 \\
 \hline
 
 \end{tabular}
 \caption{Fitting parameters - Collective relaxation model}
 \label{tab:CollectiveRelaxation}
\end{table}

The variable $E_\text{s}$, which defines the activation energy for relaxation as the glass approaches its ideal state, is constrained to some extent. Defining the longest times for which drift is reported in the literature as a lower limit for $\tau_1$, we calculate a lower limit for $E_\text{s}$. In pure GST, drift was measured for more than \unit[8*10$^6$]{$s$} at \unit[300]{$K$} \textsuperscript{\cite{Gorbenko2019}} and in a dGST, similar to the one used in this study, drift for \unit[10$^4$]{$s$} at \unit[420]{$K$} \textsuperscript{\cite{LeGallo2018}} has been reported. This corresponds to a lower limit of \unit[0.95]{$eV$} and \unit[1.12]{$eV$} for GST and dGST, respectively. The upper limit of $E_\text{s}$ is on the order of the activation energy of crystallization, which is \unit[3.2]{$eV$} for GST \textsuperscript{\cite{Jeyasingh2014a}} and \unit[3.01]{$eV$} for dGST.\textsuperscript{\cite{Sebastian2014a}} 

At this point it is worth highlighting the simplicity of this model, which has been shown to not only capture relaxation at a constant ambient temperature but also the effect of annealing profiles on the resistance in phase change memory devices.\textsuperscript{\cite{Sebastian2015}} Only two variables, $\nu_0\Delta_\Sigma$, and $E_\text{s}$ define the relaxation dynamics and, since relaxation is abstracted to a collective process, a single variable suffices to describe the degree of relaxation of the glass state.

\subsection*{Gibbs model}

\noindent The relaxation model introduced by Gibbs in the 1980s defines the glass by a spectrum of defect states with different activation energies ($q(E_\text{d})$).\textsuperscript{\cite{Gibbs1983}} The sum of these defect states ($Q$) and their change with time is the equivalent of $\Sigma(t)$ in the collective relaxation model. While the physical picture of the relaxation process is different, the models are mathematically quite similar. Like in the collective relaxation model, a rate equation with an Arrhenius dependence on the defect state activation energy for relaxation ($E_\text{d}$) is assumed
\begin{equation}
\label{equ:DifferentialGibbs}
 \frac{dq(E_\text{d},t)}{dt} = -\nu_0 * \exp\left(\frac{-E_\text{d}}{k_b T}\right) * q(E_\text{d},t)
\end{equation}
where $\nu_0$ is the attempt to relax frequency. Since the probability to relax depends exponentially on the activation energy for relaxation, defects with a small activation energy relax first and only defects in a narrow range of activation energies relax at the same time (see Supplementary Note 5). Thus, the activation energy that must be overcome for further relaxation effectively increases monotonically, like in the collective relaxation model. For a constant ambient temperature, the relaxation dynamics are given by the equation

\begin{equation}
\label{equ:GibbsConstantTamb}
 Q(t,T) = \int q(E_\text{d}) * \exp\left(-t * \nu_0 * \exp\left[\frac{-E_\text{d}}{k_\text{b}T}\right]\right) dE_\text{d}
\end{equation}

\begin{figure}
\centering
 \includegraphics[width=1\linewidth]{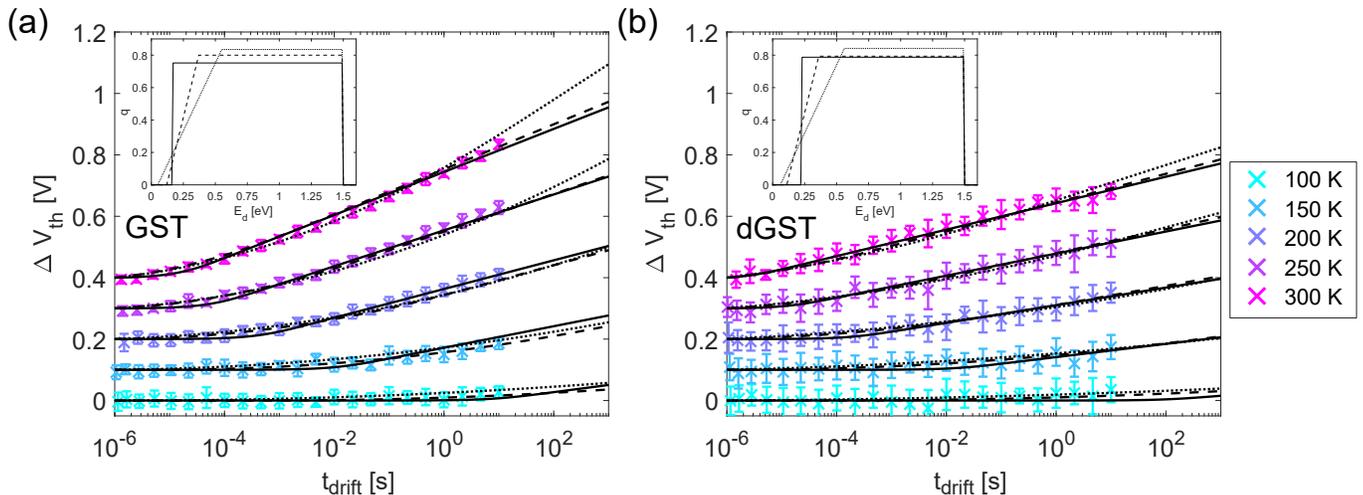}
 \caption{Model fit - Gibbs model: The temporal evolution of the threshold voltage of GST (a) and dGST (b) is fitted to the Gibbs model for three differently shaped activation energy spectra. The initial defect distribution functions $q_0(E_\text{d})$ are shown in the figure inset. When the activation energy spectrum increases over an energy range that is too wide, here \unit[0.5]{$eV$}, the onset gets stretched out too much. For better visibility the experiments at different temperatures are shifted along the y-axis by $(T_{amb}-100 K) * 0.2 V/K$. The fitting parameters are summarized in Table \ref{tab:GibbsModel}.}
 \label{fig:GibbsModel_Fit}
\end{figure}

The main challenge in probing the Gibbs model is to infer what spectrum of defect states the material has. Experimental studies on other materials show bell-shaped or more complex distribution functions.\textsuperscript{\cite{Shin1993,Chen1976,Khonik2008,Tsyplakov2014,Friedrichs1989}} To capture the strict $\log(t)$ dependence observed in phase change materials over many orders of magnitude in time, a rather flat $q(E_\text{d})$ is required.\textsuperscript{\cite{Knoll2009, LeGallo2018, Ielmini2007}} The onset of relaxation emerges from a transition from no defects to existing defects. How sharp it is depends on the width of this transition and the attempt to relax frequency. We fit the experiment assuming three different initial defect distributions $q_0(E_\text{d})$. One with a step-like transition and two with a linear transition over an energy range of \unit[0.25]{$eV$} and \unit[0.5]{$eV$} (Figure \ref{fig:GibbsModel_Fit}). The upper limit of $q_0(E_\text{d})$ is set to \unit[1.5]{$eV$}, which is well beyond the highest activation energies that can be overcome on the timescales and temperatures probed in our study. The position of the transition, the attempt to relax frequency and the proportionality constant $C_1$ are free fitting parameters. All three distributions give a fairly good fit to the experimental data. An extrapolation to longer timescales, however, shows that the onset is stretched out too much for the \unit[0.5]{$eV$} wide transition. In both materials, the $q(E_\text{d})$ needs to have a rather sharp transition from zero to a constant number of defects. 

\begin{table}[h!]
 \centering
 \begin{tabular}{ c c c c}
 \hline
 Material & Transition q(E) & $\nu_0 [s^{-1}]$ &  $C_1 [V]$ \\
 \hline
 GST & step & 3.66e7 & -1.57 \\
 GST & 0.25 eV & 7.33e8 & -1.54 \\
 GST & 0.5 eV & 8.64e7 & -2.33 \\
 dGST & step & 2.27e9 & -0.92 \\
 dGST & 0.25 eV & 1.97e9 & -1.00 \\
 dGST & 0.5 eV & 9.33e9 & -1.15 \\
 \hline
 \end{tabular}
 \caption{Fitting parameters - Gibbs model: The threshold voltage drift is fitted with three differently shaped initial activation energy spectra $q_0(E_\text{d})$ (inset - Figure \ref{fig:GibbsModel_Fit}). Dependent on the spectrum the attempt to relax frequency and the proportionality constant between $V_{th}$ and the sum of defect states changes.}
 \label{tab:GibbsModel}
\end{table}

\section{Discussion}

\noindent The onset of relaxation constrains the parameters in both relaxation models. The activation energy that must be overcome for the first relaxation step in the collective relaxation model is \unit[0.19]{$eV$} for GST and \unit[0.24]{$eV$} for dGST. In the Gibbs model, the equivalent to this is the position of the transition from no defects to existing defects, which is at around \unit[0.1]{$eV$} to \unit[0.25]{$eV$} for both materials (Figure \ref{fig:GibbsModel_Fit}, inset). It changes slightly depending on the assumed shape of $q(E_\text{d})$. In the phase change memory cell, a melt-quenched state is created with extremely high cooling rates on the order of \unit[$\sim$10$^{10}$]{$K/s$}. It thus allows studying the relaxation from an extremely unrelaxed glass state, which manifests in the low activation energy for relaxation and consequently an onset of relaxation at short time scales. 

Another parameter used in both models is the attempt to relax frequency $\nu_0$ (Equation \ref{equ:DifferentialSigma} and Equation \ref{equ:DifferentialGibbs}). Previous studies on phase change materials estimated the attempt to relax frequency in the typical phonon frequency range of 10$^{13}$ to \unit[10$^{14}$]{$s^{-1}$}.\textsuperscript{\cite{Ielmini2008a, LeGallo2018}} Our fits to the Gibbs model give an attempt to relax frequency on the order of \unit[10$^7$]{$s^{-1}$} to \unit[10$^8$]{$s^{-1}$} for GST and \unit[10$^9$]{$s^{-1}$} for dGST, which is notably lower than previously considered for phase change materials. 
In order to fit experimentally obtained relaxation dynamics to the Gibbs model, the attempt to relax frequency is commonly used as a free fitting parameter. For metallic glasses, attempt to relax frequencies ranging from 10$^{11}$ to \unit[10$^{15}$]{$s^{-1}$} have been reported. Here, reduced attempt to relax frequencies were ascribed to relaxation processes involving groups of atoms.\textsuperscript{\cite{Friedrichs1989}} In carbon doped amorphous silicon frequencies as low as \unit[10$^6$]{$s^{-1}$} have been found.\textsuperscript{\cite{Stutzmann1986}} To fit the relaxation of nanoscale indents in a polymer glass an attempt to relax frequency of \unit[2*10$^{24}$]{$s^{-1}$} has been used.\textsuperscript{\cite{Roura2009}} We believe that to justify the Gibbs model, further explanation, and physical reasoning, as to why the attempt to relax frequency could change over so many orders of magnitude, is required. The corresponding fitting parameter in the collective relaxation model is $\nu_0\Delta_\Sigma$ which is \unit[$\sim$10$^6$]{$s^{-1}$} for GST and \unit[$\sim$10$^8$]{$s^{-1}$} for dGST. The variable $\Delta_\Sigma$ is expected to be $<<1$; $1/\Delta_\Sigma$ is the hypothetical number of the different configurational states that the system could assume. Accordingly, the collective relaxation model requires orders of magnitude higher attempt to relax frequencies than the Gibbs model. In this case we can expect attempt to relax frequencies of 10$^{13}$ to \unit[10$^{14}$]{$s^{-1}$}.

A major critique against the Gibbs model concerns the shape of the activation energy spectrum required to capture the drift of phase change materials. Studies on metallic glasses found bell shape or more complex spectra with a rather shallow increase of the number of defect states over a range of 0.5 to \unit[1]{$eV$}.\textsuperscript{\cite{Khonik2008,Tsyplakov2014,Friedrichs1989,Chen1976}} Opposed to this, to explain the  $\log(t)$ dependent drift observed over many orders of magnitude in time a rather flat spectrum of defect states over a range of at least \unit[1]{$eV$} is required.\textsuperscript{\cite{LeGallo2018}} Additionally, to capture the relaxation onset, the transition between no defects to existing defects must happen in a narrow energy range. The threshold voltage drift characterized here represents an average behavior of multiple RESET states created in the device. Thus, the relaxation onset is blurred, and the characterization of a single glass state would probably show an even sharper relaxation onset. A sharp onset also requires a sharp transition of $q(E_\text{d})$. These considerations indicate that an almost step-like $q(E_\text{d})$ is required to capture the relaxation onset in phase change materials. This provides further evidence that an improbable $q(E_\text{d})$ is required to explain the drift of phase change materials with the Gibbs model. In fact, for the scenario of a step-like transition, the Gibbs model and the collective relaxation model give identical fits to our experimental data (Supplementary Note~5). To constrain $q(E_\text{d})$ further, relaxation studies over even longer timescales are required. 

Two recently proposed relaxation models postulate that resistance drift may also result from the release of trapped electrons. The release of these charge carriers has been proposed to increase the width of the potential barrier needed to overcome at the contact between electrode and phase change material.\textsuperscript{\cite{Khan2020}} A second hypothesis states that these electrons recombine with thermally generated holes in the valence band and thus reduce the number of free charge carriers in the amorphous state.\textsuperscript{\cite{Elliott2020}} Even though these models assume quite a different mechanism, they could in principle explain the onset and saturation of drift. The onset of drift would be determined by the potential barrier and attempt to escape frequency of electrons from a trap state. Drift would saturate when an equilibrium between electron trapping and detrapping is reached. In the current version, however, the model proposed in \cite{Elliott2020} is designed such that drift begins immediately after RESET. Neither of the two models specifies the expected dynamics of electron detrapping. Thus, we cannot say if or how well these models will be capable of quantitatively capturing the temperature-dependent onset of relaxation. 

\section{Conclusion}
\noindent In this work we experimentally measured the onset of structural relaxation in melt-quenched amorphous phase-change materials. Threshold-switching voltage was used to measure the state of relaxation. Experiments were performed using mushroom-type phase-change memory devices with GST and doped GST as phase-change materials. The onset of structural relaxation, marked by a transition from almost constant threshold-switching voltage values to the commonly observed $\log(t)$ dependence, changes profoundly with ambient temperature; from microseconds at \unit[300]{$K$} to tens of seconds at \unit[100]{$K$}. We found that both the Gibbs relaxation model and the collective relaxation model are capable of describing the experimental data. The fits to the Gibbs model, however, required an almost step-like defect distribution and orders of magnitude lower attempt to relax frequencies than estimated in previous works. 

\section{Methods}
\noindent \textit{Mushroom-type PCM device:}
 The mushroom-type PCM cells used in this study were fabricated in the 90-nm technology node. The multi-layer ring bottom electrode with a radius of \unit[$\sim$20]{$nm$} and a height of \unit[$\sim$40]{$nm$} was patterned with a sub-lithographic hardmask process. The sputter deposited phase change material is \unit[$\sim$75]{$nm$} thick. To assure a stable device operation throughout our study the cells were cycled at least 100,000 times in advance. The devices are fabricated with an on-chip series resistor of \unit[$\sim$2]{$k\Omega$}. 

\noindent \textit{Experimental setup:}
The experiments were performed in a cryogenic probe station (JANIS ST-500-2-UHT), cooled with liquid nitrogen, that operates between 77 to \unit[400]{$K$}. The sample holder and chamber temperature was controlled with an accuracy of \unit[$\pm$0.5]{$K$}. 
AC voltage signals were applied to the device with an Agilent 81150A Pulse Function Arbitrary Generator. To send the SET pulse (Instrument Output 1) with a defined delay time after the RESET pulse (Instrument Output 2), the two instrument outputs were coupled internally. Cell voltage and current were measured with a Tektronix DPO5104B digital oscilloscope, which was triggered on the SET pulse leading edge. The transient signals were sampled with a frequency of \unit[2.5]{$GHz$}.

\noindent \textit{Experimental protocol:}
In order to measure the threshold voltage evolution with time the device is programmed to a new RESET state multiple times  and the pause t\textsubscript{delay} before applying the SET pulse is increased. This sequence with a t\textsubscript{delay} ranging from \unit[10]{$ns$} to \unit[10]{$s$} is repeated 15 times to average out drift and threshold switching variability. Due to the variability some scattering of the data and potentially a blurring of the onset of relaxation is inevitable. Nonetheless the overall threshold voltage change with time is a smooth curve. The standard deviation is around \unit[30]{$mV$} for GST and \unit[50]{$mV$} for dGST.

\noindent \textit{Definition of the threshold voltage:}
To extract the threshold voltage from the switching IV curve we fit the load-line of the voltage snap-back (Supplementary Note 6). The mushroom cell is fabricated with an on-chip series resistor. In the moment of switching the cell resistance drops to values similar to the series resistor (R\textsubscript{ser}) and thus the voltage drop over the cell decreases. By fitting the load-line, instead of choosing the largest voltage drop prior to switching as threshold voltage value, the analysis scheme becomes more resilient to noise in the transient voltage and current trace. The threshold voltage is defined at a load-line current of \unit[5]{$\mu A$}.

\noindent \textit{Impact of the SET pulse shape on \Vth}
To induce threshold switching the device is biased with a triangular voltage pulse. Both, the electrical stress and joule heating in the device prior to switching, may affect the relaxation dynamics of the device. In fact, the threshold voltage of a nanoscale device changes dependent on the transient voltage signal applied in order to switch the device.\textsuperscript{\cite{Wimmer2014a,LeGallo2016}} With an increasing duration of the SET pulse leading edge, the threshold voltage decreases (Supplementary Figure 8a). The absolute change of $V_{th}$ with time, however, appears to be independent of the leading edge (Supplementary Figure 8b). This suggests that the rise of the cell bias is fast enough to not notably alter the relaxation process of the glass state prior to switching. First, the time for which the cell is biased prior to switching is short on absolute time scales. Second, it is at least in regime 2 and 3, which are governed by the relaxation dynamics, much shorter than the time for which the material relaxes without any bias being applied.



\section*{Conflict of Interest}
The authors declare no conflict of interest.

\section*{Acknowledgements}
This work was supported by the IBM Research
AI Hardware Center. This work was also partially funded by the European Research Council (ERC) under the European Union’s Horizon 2020 research and innovation program (grant agreement numbers 682675 and 640003).

\beginsupplement

\newpage

\huge\textbf{Supplementary Notes}

\normalsize
\section{Comparison to prior threshold voltage drift experiments} 

Careful inspection of the threshold voltage drift experiment at room temperature reported in \cite{Ielmini2007} shows the same three regimes we observe in our study. While a continuous drift with log(t) is observed from about \unit[$10^{-5}$]{$s$} in both studies, regime 1 occurs at different timescales. In our study it lasts up to \unit[1]{$\mu s$} and in the work by Ielmini et al. it ends after about \unit[30]{$ns$}. This could be attributed to the devices having different geometries and sizes (mushroom cell with a bottom electrode \unit[$\sim$ 706]{$nm^2$} vs. $\mu$-trench with a bottom electrode \unit[>1500]{$nm^2$}) and the resulting difference in the thermal environment. Additionally, in \cite{Ielmini2007} the bias of the RESET pulse is reduced to a lower level, sufficient to melt-quench and induce threshold switching, but it is not turned off between RESET and threshold switching. The experiment probes the decay time to a partially excited state ($I \sim 150 \mu A$) defined by the bias applied at the end of the RESET pulse. Thus, the different bias scheme compared to our experiment, where the bias is completely switched off at the end of the RESET pulse, is another reason why regime 1 lasts longer in our study.

\section{Threshold voltage transients in regime 1} 

\begin{figure}[h!]
	\centering
	\includegraphics[width=0.8\linewidth]{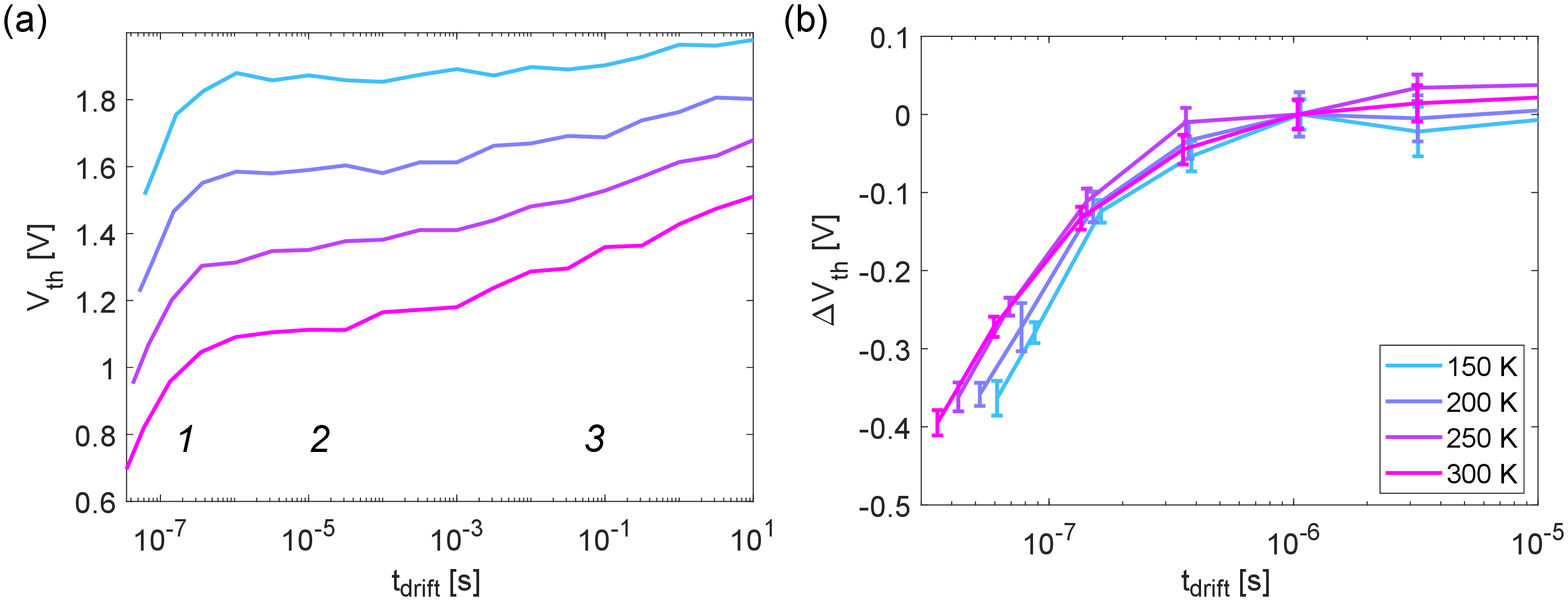}
	\caption{Threshold voltage evolution: (a) The threshold voltage evolution shows the same three characteristic regimes (labeled in the figure) for all temperatures studied. At lower ambient temperatures the first measurement point is delayed because more time elapses before the applied voltage pulse with a \unit[100]{$ns$} leading edge reaches $V_{th}$. (b) The threshold voltage change in regime 1 is calculated with respect to the value measured 1 $\mu s$ after RESET. In regime 1 the temporal evolution appears to be independent of the ambient temperature.}
	\label{fig:DriftRegime1}
\end{figure}

The threshold voltage evolution with time shows three distinct regimes (Figure \ref{fig:DriftRegime1} a). While the transition between regimes 2 and 3, which marks the onset of relaxation, and the drift coefficient in regime 3 exhibit a pronounced temperature dependence, regime 1 appears to be almost independent of the ambient temperature (Figure \ref{fig:DriftRegime1} b). Between \unit[30]{$ns$} and \unit[1]{$\mu s$} after RESET the threshold voltage increases by \unit[0.4]{$V$}. This rapid change is most likely caused by the decay of the RESET excitation. On one hand, a large number of excess charge carriers is generated when the device is molten at high-fields, on the other hand the temperature is locally raised to more than \unit[900]{$K$} before the applied power is switched off within \unit[3]{$ns$} (RESET pulse trailing edge). Both a decay of the excess charge carriers \cite{Elliott2020,Ielmini2007} and a slow decay of the local temperature when the device is already close to ambient temperature would result in a continuous increase of the threshold voltage.  

\newpage

\section{RESET state programming} 

To create comparable RESET states at different ambient temperatures, the programming power was scaled such that the initially molten volume remained approximately constant. The applied quench-rate, defined by the \unit[3]{$ns$} RESET pulse trailing edge, was kept constant. An easily accessible metric to compare the molten volume at different ambient temperatures is the hot-spot temperature ($T_{hs}$) inside the device. If $T_{hs} = R_{th} P_{inp} +T_{amb}$ remains constant, so does the molten volume. $R_{th}$ is the average thermal resistance of the device, $P_{inp}$ the input power associated with the voltage pulse and $T_{amb}$ the ambient temperature.\textsuperscript{\cite{Boniardi2012, Sebastian2014a}} 

\begin{figure}[h!]
	\centering
	\includegraphics[width=0.9\linewidth]{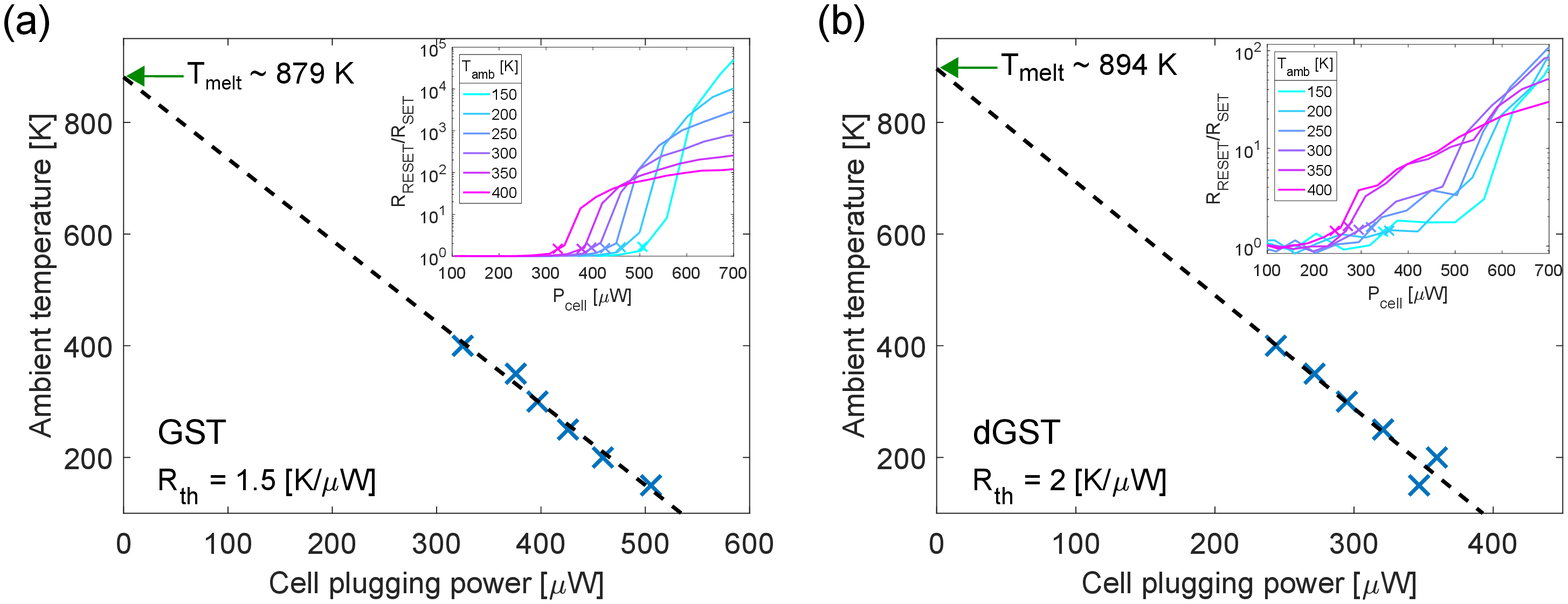}
	\caption{Thermal resistance: Programming curves of (a) GST and (b) dGST cells are measured at ambient temperatures ranging from \unit[150]{$K$} to \unit[400]{$K$} (insets). The cell plugging power, defined as the programming power required to induce a first increase of cell resistance, increases linearly with ambient temperature. The slope of the linear fit is the device's thermal resistance.}
	\label{fig:ThermalResistance}
\end{figure}

The thermal resistance of both devices is obtained from the programming curves measured at different ambient temperatures (Figure \ref{fig:ThermalResistance}). The programming power at which the device resistance begins to increase marks the point at which $T_{hs}$ rises to the melting temperature $T_{melt}$ and an amorphous volume begins to cover the device bottom electrode. Extrapolated to \unit[0]{$\mu W$}, the temperature-dependent programming power coincides fairly well with the melting temperature. GST has a melting temperature of \unit[858]{$K$} \textsuperscript{\cite{Adnane2017}} and for a comparable type of doped GST a melting temperature of \unit[877]{$K$} was measured.\textsuperscript{
	\cite{Sebastian2014a}} 

\begin{figure}[h!]
	\centering
	\includegraphics[width=0.45\linewidth]{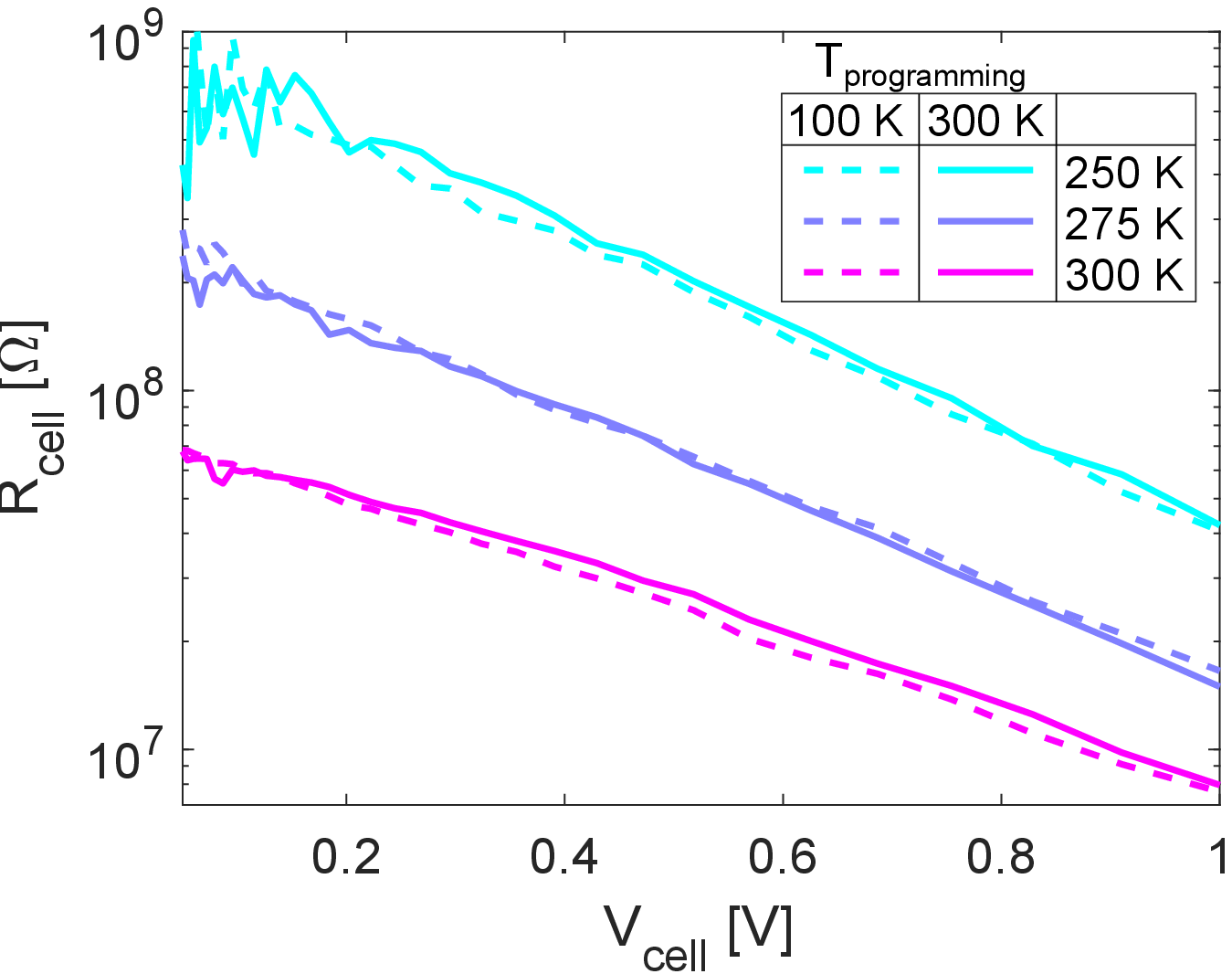}
	\caption{Programming comparable device states: Two devices are programmed with the same $T_{hs}$ at an ambient temperature of \unit[100]{$K$} and \unit[300]{$K$}. After annealing, the RV characteristic of the device state is probed at ambient temperatures from \unit[300]{$K$} to \unit[250]{$K$}. RESET states with identical transport characteristics were created in both devices. This demonstrates that amorphous volumes of similar size were programmed at \unit[100]{$K$} and \unit[300]{$K$}}.
	\label{fig:ComparableStates}
\end{figure}

A comparison of two dGST devices, which were programmed with the same $T_{hs}$ at an ambient temperature of \unit[100]{$K$} and \unit[300]{$K$}, shows the fidelity of this approach (Figure \ref{fig:ComparableStates}). The two cell states have a different thermal history and thus represent differently relaxed glass states. An annealing step to \unit[320]{$K$} for \unit[15]{$minutes$} is applied to erase the differing thermal history. After annealing the two devices show an identical field and temperature-dependent transport. This remarkable match, even though the devices were programmed at different ambient temperatures, shows two things. First, that first RESET states of comparable size were created. Second, that upon annealing the glass state in both devices exhibits identical transport characteristics, which implies that the annealing step created similarly relaxed glass states.

\newpage

\section{Collective relaxation and thermally assisted threshold switching} 
\label{Supp:FullModel}

In the main manuscript, we propose that $V_{th}$ changes approximately proportional to $\Sigma$ and that the change upon relaxation is decoupled from the change with ambient temperature
\begin{equation}
	V_{th} = f(T_{amb}) + C_1 * \Sigma(t,T) + C_2 
\end{equation}
This approximation is further supported by simulations based on previously published models. In this work, a field dependent transport model,\textsuperscript{\cite{Gallo2015}} the collective relaxation model \textsuperscript{\cite{LeGallo2018}} and a threshold switching model \textsuperscript{\cite{LeGallo2016}} are combined. Here the models are briefly sketched to introduce the most relevant variables and assumptions. 

The electrical transport in amorphous phase change materials is highly field dependent. It can be modeled as a multiple-trapping transport together with 3D Poole-Frenkel emission from a two-center Coulomb potential.\textsuperscript{\cite{Gallo2015}} At low fields, the density of free charge carriers depends on the activation energy for conduction $E_a$, which corresponds to the  depth of the coulomb potential. The activation energy follows a Varshni law temperature dependence $E_a = E_{a0} - \frac{aT^2}{b+T}$.\textsuperscript{\cite{Varshni1967}} With increasing field strength, the Coulomb potentials of neighboring defect states, separated by the intertrap distance $s$, overlap and the effective barrier height for emission thus decreases.

Structural relaxation results in an increase of the activation energy and the intertrap distance. The activation energy changes proportionally to the state variable of the glass $E_{a0} = E^* - \alpha \Sigma(t)$ and the intertrap distance with $s(t) = s_0  /\Sigma(t)$.\textsuperscript{\cite{LeGallo2018}}

In a nanoscopic device structure, threshold switching can be induced by a thermal feedback loop.\textsuperscript{\cite{LeGallo2016}} The device is described as a thermal RC-circuit with the electrical power applied to the device as input variable. The Joule heating at elevated fields increases the temperature, which in turn increases the density of free charge carriers and thus the Joule heating.

\begin{figure}[h!]
	\centering
	\includegraphics[width=0.8\linewidth]{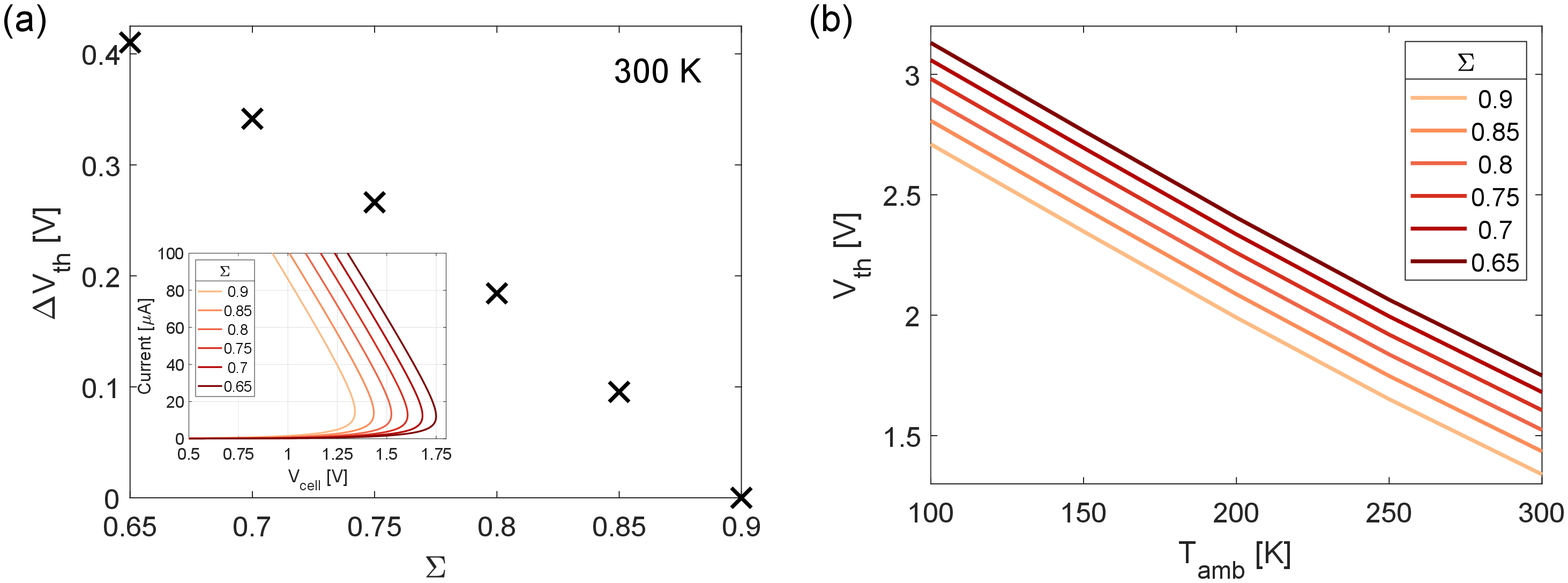}
	\caption{Threshold voltage modeling: (a) The threshold voltage changes approximately linearly with relaxation. The inset shows the simulated threshold switching IV characteristic. (b) The temperature dependence of $V_{th}$ shows no notable change with $\Sigma$.}
	\label{fig:SigmaVth}
\end{figure}

To simulate the temperature dependence of $V_{th}$ for differently relaxed glass states, the three models are combined and solved numerically. The threshold switching dynamics are simulated for a \unit[3.5]{$V$} pulse with a \unit[500]{$ns$} leading edge. All model parameters are summarized in Table \ref{tab:SwitchingSimulations}. The threshold voltage change is calculated with respect to an initial glass state $\Sigma_0 = 0.9$ and the range of $\Sigma$ is defined such that $\Delta V_{th}$ matches the experimentally observed threshold voltage increase. The simulation results show that a linear dependence of $V_{th}$ on $\Sigma$ is a good approximation (Figure \ref{fig:SigmaVth} a). 
Additionally, differently relaxed glass states show a similar change of $V_{th}$ with ambient temperature (Figure \ref{fig:SigmaVth} b). The calculated $V_{th}$ values follow parallel lines. This supports the assumption that the change with temperature is decoupled from the change upon relaxation. 

\begin{table}[h!]
	\centering
	\begin{tabular}{|l|l|}
		\multicolumn{2}{c}{\textbf{Transport model}} \\ \hline
		$r_{BE}$ [nm] & 32 \\
		$u_a$ [nm] & 10.94 \\ 
		$\epsilon_r$ & 10 \\ 
		$K*\mu_0 [m^{-1}V^{-1}s^{-1}]$ & $10^{22}$ \\ 
		a [$\mu eV K^-1$] & 600 \\ 
		b [K] & 800 \\ \hline
		\multicolumn{2}{c}{\textbf{Relaxation model}} \\ \hline
		$E^*$ [eV]& 0.505 \\
		$\alpha$ [eV] & 0.267 \\
		$s_0$ [nm] & 2.08 \\ \hline
		
		\multicolumn{2}{c}{\textbf{Thermal model}} \\ \hline
		$R_{th} [K / \mu W] $ & 3.7 \\
		$\tau_{th} = R_{th} * C_{th} [ns] $ & 4.1 \\
		$R_{ser} [\Omega]$ & 6170 \\ \hline
	\end{tabular}
	\caption{Model parameters to simulate the threshold switching: The amorphous volume in the mushroom cell is approximated as a cylinder with radius $r_{BE}$ and height $u_a$. $\epsilon_r$ denotes the relative high-frequency dielectric constant and $K*\mu_0$ a model constant. a and b are the parameters of the Varshni law. The parameter $\alpha$ is taken from.\textsuperscript{\cite{LeGallo2018}} All other variables are from.\textsuperscript{\cite{LeGallo2016}} $E^*$ and $s_0$ are defined such that $E_{a0}$ and $s$ correspond to the values reported in \cite{LeGallo2016} for $\Sigma = 0.8$.  $R_{ser}$ is the on-chip resistor in series with the phase change memory cell.}
	\label{tab:SwitchingSimulations}
\end{table}

\newpage

\section{Relaxation models comparison} 

The basic idea of the Gibbs model is that the glass state can be described by a distribution of defect states. These defects relax individually, without creating new defects or changing the activation energy for the relaxation of defect states in their local surrounding. The collective relaxation model on the other hand describes relaxation as a sequence of transitions between neighboring unrelaxed configurational states. Upon relaxation local configurations become more stabilized but are still involved in subsequent collective rearrangements. Thus, the activation energy for relaxation increases with each relaxation step. 

\begin{figure}[h!]
	\centering
	\includegraphics[width=1\linewidth]{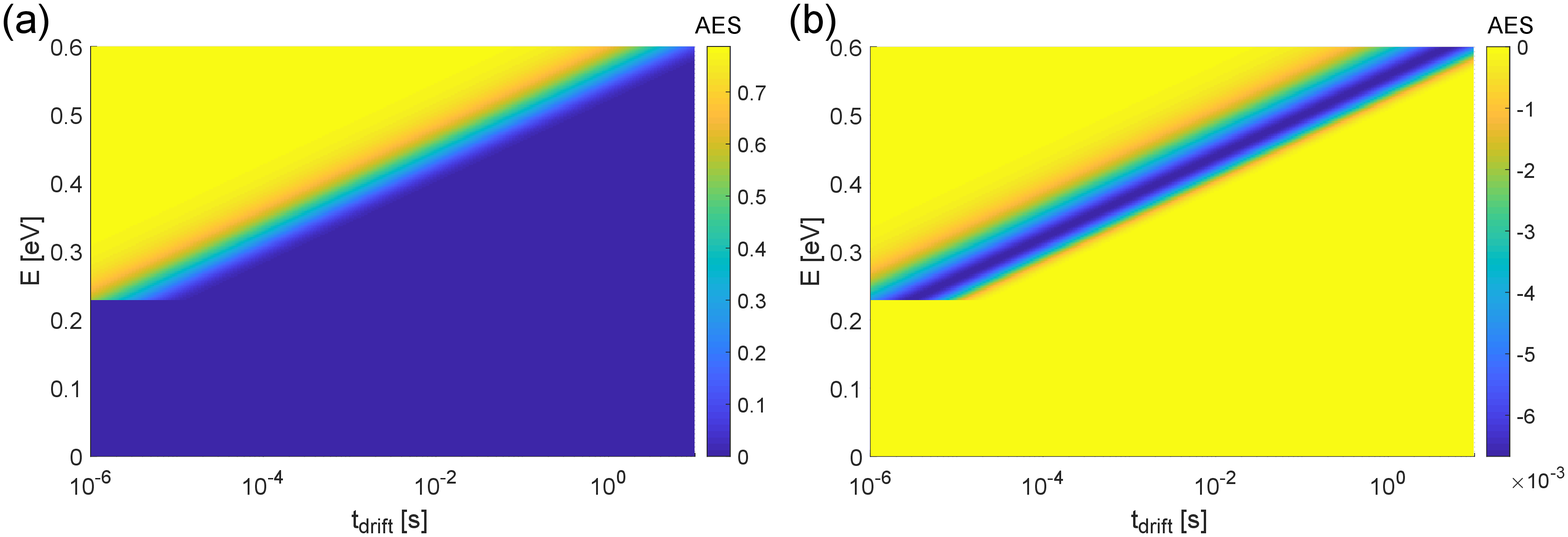}
	\caption{dGST activation energy spectrum: The spectrum is derived from the fit in Figure \ref{fig:CollectiveVsGibbs} b, assuming a flat defect distribution. (a) Evolution of the activation energy spectrum upon relaxation at \unit[300]{$K$}. The lower limit of the spectrum increases linearly with log(t). (b) Change of the activation energy spectrum with time. At each time instance defect states within a narrow range of activation energies relax.}
	\label{fig:AESdrift}
\end{figure}

\begin{figure}[h!]
	\centering
	\includegraphics[width=1\linewidth]{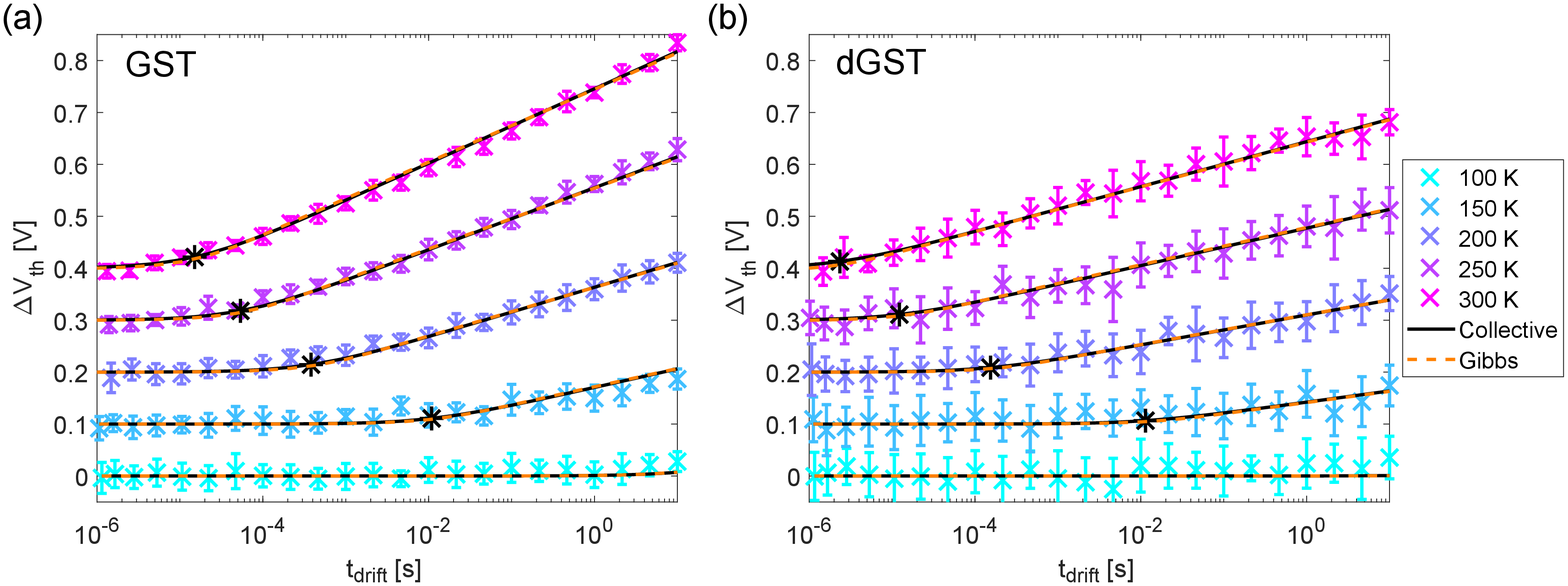}
	\caption{Model comparison: A one-on-one comparison of the collective relaxation model and the Gibbs model for the scenario of a flat $q(E_\text{d})$ with a step-like transition from no defects to existing defects shows almost identical fits. The fitting parameters are summarized in table 1 and table 2 of the main manuscript.}
	\label{fig:CollectiveVsGibbs}
\end{figure}

Despite different concepts of the relaxation process, the Gibbs model 
\begin{equation}
	\frac{dq(E_\text{d},t)}{dt} = -\nu_0 * exp(\frac{-E_\text{d}}{k_b T}) * q(E_\text{d},t)
\end{equation}
and the collective relaxation model 
\begin{equation}
	\frac{d\Sigma(t))}{dt}= - \nu_0 \Delta_\Sigma *\exp(\frac{E_\text{s}*(1-\Sigma(t))}{k_\text{b}T})
\end{equation}
are mathematically constructed similar, in the sense that both assume a first order rate equation with an Arrhenius temperature dependence on the activation energy of relaxing defects. In the collective relaxation model, at any time instance only one activation energy governs the next relaxation step, whereas in the Gibbs model multiple defects with different activation energies can relax simultaneously. But at each time instance, only a narrow range of activation energies has a finite probability of relaxing (Figure \ref{fig:AESdrift}). Effectively, the activation energy of relaxing defects continuously increases, like in the collective relaxation model. Assuming a flat distribution $q(E_\text{d})$, with a step like transition from no defects to existing defects for the Gibbs model, both models give almost identical fits to our experimental data (Figure \ref{fig:CollectiveVsGibbs}).

\newpage
\section{Algorithm to determine V\textsubscript{th}} 

\begin{figure}[h!]
	\centering
	\includegraphics[width=0.6\linewidth]{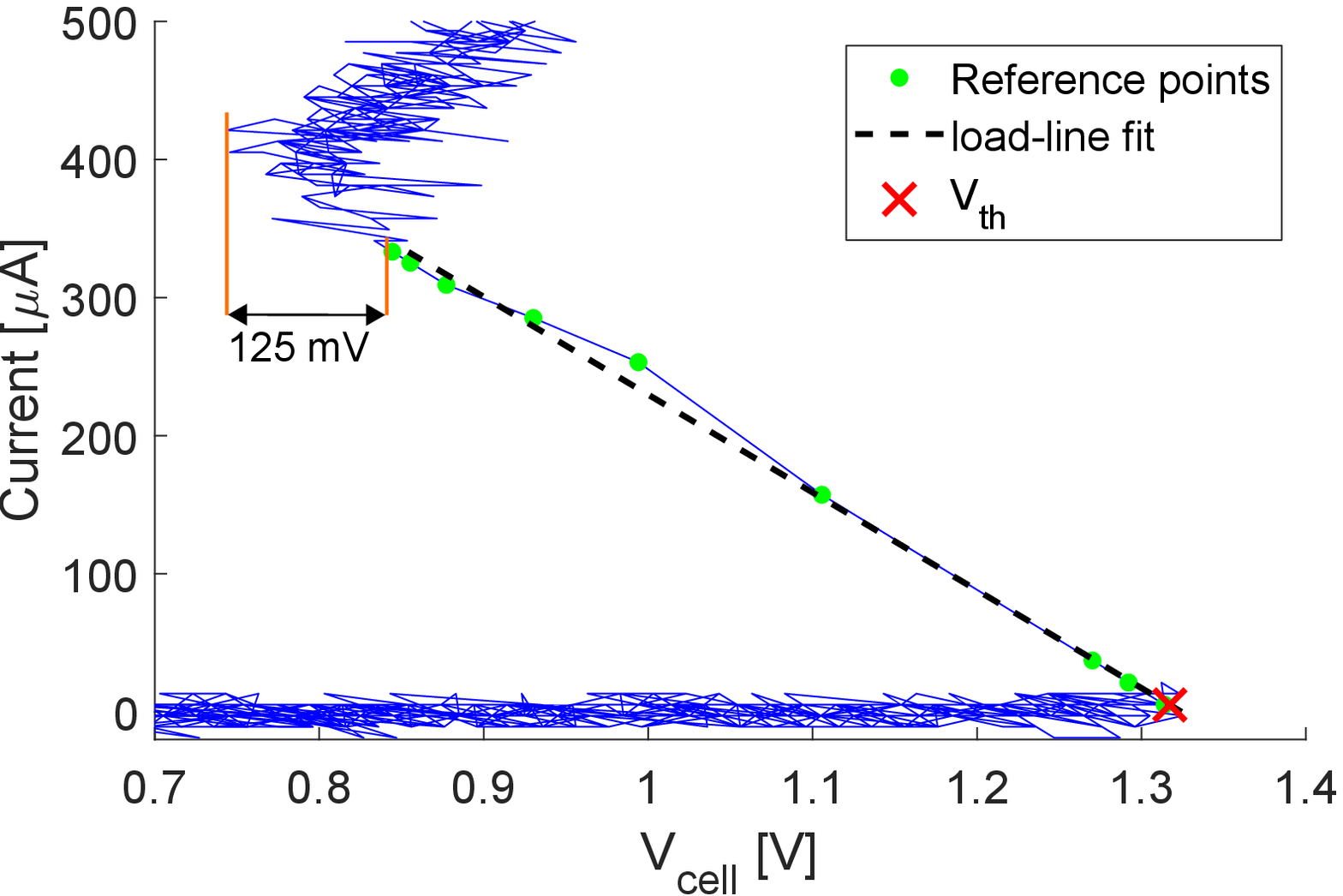}
	\caption{Threshold switching IV characteristic: The snap-back of the IV curve is fitted linearly to obtain the threshold voltage. The threshold voltage is defined at a load-line current of \unit[5]{$\mu A$}. Green dots mark the measurement points used to fit the load-line.}
	\label{fig:DefenitionVth}
\end{figure}

The threshold voltage value is obtained by fitting the load-line of the switching $IV$ curve (Methods). The reference points to fit the load-line range from the last point where the device current is smaller than \unit[20]{$\mu$A} to \unit[125]{$mV$} above the minimum voltage of the load-line (Figure \ref{fig:DefenitionVth}). This approach is taken to make the analysis scheme more resilient to noise in the transient voltage and current traces. 

\newpage
\section{Impact of the SET pulse shape} 

\begin{figure}[h!]
	\centering
	\includegraphics[width=0.9\linewidth]{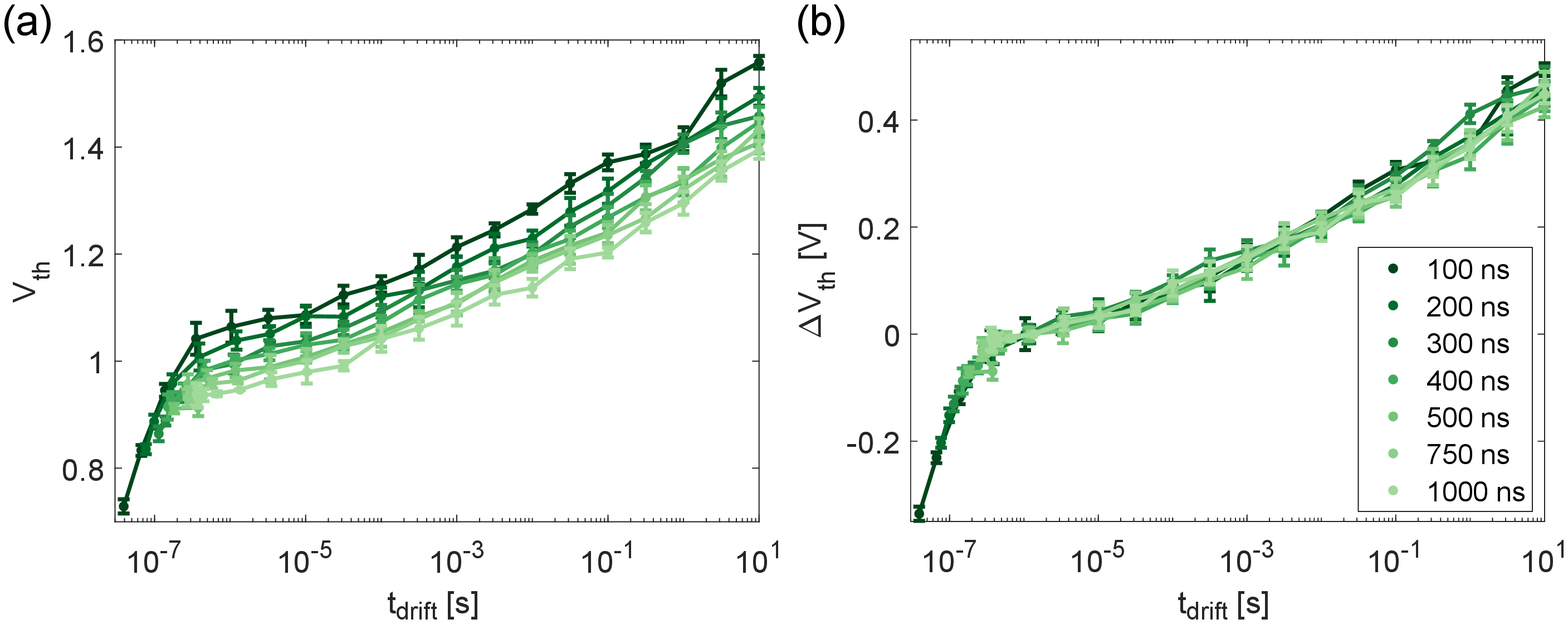}
	\caption{Impact of the SET pulse leading edge duration: (a) With an increasing SET pulse leading edge the threshold voltage values decrease continuously. The value of $t\textsubscript{drift}$ is the sum of $t\textsubscript{delay}$ (the delay between the RESET pulse and the SET pulse) and the time elapsed until the applied voltage crosses $V\textsubscript{th}$. The second term changes with the SET pulse leading edge. The resulting change of $t\textsubscript{drift}$ is most apparent on short timescales, where $t\textsubscript{delay}$ is small. (b) The threshold voltage change with respect to a reference point, which again is the value measured \unit[1]{$\mu s$} after RESET, is independent of the SET pulse leading edge duration. The SET pulse amplitude is \unit[2.5]{$V$}.}
	\label{fig:SET_leading_edge}
\end{figure}

To induce threshold switching, the device is biased with a triangular voltage pulse. With an increasing duration of the SET pulse leading edge, the threshold voltage decreases (Figure \ref{fig:SET_leading_edge} a). The absolute change of $V_{th}$ with time, however, appears to be independent of the leading edge (Figure \ref{fig:SET_leading_edge} b).

\newpage

\newpage
\bibliography{literature}

\end{document}